\documentstyle[epsf,11pt]{article}

\begin{document}

\title{{\bf A Directed Continuous Time Random Walk Model with Jump Length Depending on Waiting Time}}
\author{Long Shi$^{1,2}$, Zuguo Yu$^{1,3}$\thanks{Corresponding author,
e-mail:  yuzg1970@yahoo.com or yuzg@xtu.edu.cn}, Zhi Mao$^1$, Aiguo Xiao$^1$ \\
{\small $^1$Hunan Key Laboratory for Computation and Simulation in
Science and Engineering and Key }\\
{\small Laboratory of Intelligent Computing and Information
Processing of Ministry of Education,}\\
{\small Xiangtan University, Xiangtan,  Hunan 411105, China.}\\
{\small $^{2}$Institute of Mathematics and Physics, Central South
University of Forest and
Technology, Changsha,}\\
{\small  Hunan 410004, China.}\\
{\small $^3$School of Mathematical Sciences, Queensland University
of Technology, GPO Box 2434,} \\
{\small Brisbane, Q4001, Australia.} }
\date{}

\maketitle

\begin{abstract}
In continuum one-dimensional space, a coupled directed continuous
time random walk model is proposed, where the random walker jumps
toward one direction and the waiting time between jumps affects
the subsequent jump. In the proposed model, the Laplace-Laplace
transform of the probability density function $P(x,t)$ of finding
the walker at position $x$ at time $t$ is completely determined by
the Laplace transform
 of the probability density function $\varphi(t)$ of the waiting time.
 In terms of the probability density function of the waiting time
 in the Laplace domain, the limit distribution of the random process and the corresponding evolving equations are derived.
\end{abstract}

\section*{1. Introduction}

\ \ \ \  The continuous time random walk (CTRW) theory,
which was introduced by Montroll and Weiss [1] to study random walks on a lattice,
has been applied successfully in many fields (see, e.g., the reviews [2-4] and references therein).

In continuum one-dimensional space, a CTRW process is generated by
a sequence of independent identically distributed (IID) positive
waiting times $T_1, T_2, T_3, \cdots$, and a sequence of IID
random jump lengths $X_1, X_2,X_3, \cdots$. Each waiting time has
the same probability density function (PDF) $\varphi(t), t\geq0$,
and each jump length has the same PDF $\lambda(x)$ (usually chosen
to be symmetric $\lambda(x)=\lambda(-x)$). Setting $t_0=0,
t_n=T_1+T_2+\cdots +T_n$ for $n\in N$ and $x_0=0,
x_n=X_1+X_2+\cdots+X_n, x(t)=x_n$ for $t_n\leq t< t_{n+1}$, we get
a microscopic description of the diffusion process [5]. If
$\{X_n\}$ and $\{T_n\}$ are independent, the CTRW is called
decoupled. Otherwise it is called coupled CTRW [6]. The decoupled
CTRW, which is completely determined by mutually independent
random jump length and random waiting time, has been widely
studied in recent years [3-20].

In some applications it becomes important to consider coupled CTRW
[7-8]. The coupled CTRW can be described by the joint PDF
$\phi(x,t)$ of jump length and waiting time. Because
$\phi(x,t)dxdt$ is the probability of a jump to be in the interval
$(x,x+dx)$ in the time interval $(t,t+dt)$, the waiting time PDF
$\varphi(t)=\int_{-\infty}^{+\infty}\phi(x,t)dx$ and the jump
length PDF $\lambda(x)=\int_{0}^{+\infty}\phi(x,t)dt$ can be
deduced. Some kinds of couplings and correlations were proposed in
[21-25], where the symmetric jump length PDF is chosen. For the
coupled CTRW, there exist two coupled forms:
$\phi(x,t)=\lambda(x)\varphi(t|x)$ and
$\phi(x,t)=\varphi(t)\lambda(x|t)$. The first coupled form has
been studied sufficiently in many literatures [8, 21-23]. The
famous model is L\'{e}vy walk. Recently, we considered the second
coupled form, discussed the asymptotic behaviors of the coupled
jump probability density function in the Fourier-Laplace domain,
and derived the corresponding fractional diffusion equations from
the given asymptotic behaviors [25].

In this work, we introduce a directed CTRW model with jump length
depending on waiting time (i.e. $\phi(x,t)=\varphi(t)\lambda(x|t),
x>0, t>0$). In our model, the Laplace-Laplace transform [26] of
$P(x,t)$ of finding the walker at position $x$ at time $t$ is
completely determined by the Laplace transform of $\varphi(t)$.
Generally, CTRW processes can be categorised by the mean waiting
time $T=\int_{0}^{+\infty}t\varphi(t)dt$ being finite or infinite.
Here we find that the long-time limit distributions of the PDF
$P(x,t)$ are a Dirac delta function for finite $T$ and a beta-like
density for infinite $T$, the corresponding evolving equations are
a standard advection equation for finite $T$ and a
pseudo-differential equation with fractional power of coupled
space and time derivative for infinite $T$.

This paper is organized as follows. In section 2, we introduce the basic concepts of the coupled CTRW.
In section 3, a coupled directed CTRW model is introduced. In section 4,
the limit distributions and the corresponding evolving equations of the coupled directed CTRW model are derived.
The conclusions are given in section 5.

\section*{2. The coupled continuous time random walk}

\ \ \ \  Now we recall briefly the general theory of CTRW [3]. Let
$\eta(x,t)$ is the PDF of just having arrived at position $x$ at
time $t$. It can be expressed by $\eta(x',t')$ (the PDF of just
having arrived at position $x'$ at time $t'<t$) as:
\begin{equation}
\label{(1)}
\eta(x,t)=\int_{-\infty}^{+\infty}dx'\int_0^{+\infty}dt'\eta(x',t')
\phi(x-x',t-t')+\delta(x)\delta(t).
\end{equation}
Then, the PDF $P(x,t)$ with the initial condition $P(x,0)=\delta(x)$ can be described by the following integral equation [3]
\begin{equation}
\label{(2)}
P(x,t)=\int_0^t\eta(x,t')\omega(t-t')dt',
\end{equation}
where $\omega(t)=1-\int_0^t\varphi(\tau)d\tau$ is the probability of
not having made a jump until time $t$.

Let $\widehat{f}(k)$ and $\widetilde{g}(s)$ be the transforms of
Fourier and Laplace of sufficiently well-behaved (generalized)
functions $f(x)$ and $g(t)$ respectively, defined by
\begin{equation}
\label{(3)}
\widehat{f}(k)=
{\cal F}\{f(x);k\}
=\int_{-\infty}^{+\infty}f(x)e^{ikx}dx,\hspace{0.5cm} k\in R,
\end{equation}

\begin{equation}
\label{(4)}
\widetilde{g}(s)=
{\cal L}\{g(t);s\}
=\int_0^{+\infty}g(t)e^{-st}dt,\hspace{0.5cm} s>s_0.
\end{equation}

After using the Fourier-Laplace transforms and the convolution
theorems for integral equation (2), one can obtain the following
famous algebraic relation [3]
\begin{equation}
\label{(5)}
\widehat{\widetilde{P}}(k,s)=\frac{1-\widetilde{\varphi}(s)}{s}\cdot\frac{1}{1-\widehat{\widetilde{\phi}}(k,s)}.
\end{equation}

\section*{3. A coupled directed CTRW model}

\ \ \ \ In Ref. [23], the author considered a CTRW model with
waiting time depending on the preceding jump length, where the
author supposed that the PDF of the waiting time is a function of
a preceding jump length. In that model, the author introduced a
natural "physiological" analogy: after making a jump one needs
time to rest and recover. The longer the jump distance is, the
recovery and the waiting time needed are longer. This is an
interesting hypothetical physiological example. Motivated by this,
we consider a directed CTRW model with jump length depending on
the waiting time and give an analogue physiological explanation.

A directed CTRW model with jump length depending on the waiting
time can be generated by a sequence of IID positive waiting times
$T_1, T_2, T_3, \cdots$, and a sequence of jumps $X_1, X_2,
X_3,\cdots$. each waiting time has the same PDF $\varphi(t), t\geq
0$. Every time jump has the same direction and each jump length
has the same conditional PDF $\lambda(x|t), x\geq 0$, which is the
PDF of the random walker making a jump of length $x$ following a
waiting time $t$.

A natural assumption
is that the jump length is proportional to the waiting time. So we can take the simplest jump
length PDF as $\lambda(x|t)=\delta(x-vt), v>0$. Without the loss of generality, we take $v=1$ in
the following discussion. Setting $t_0=0, t_n=T_1+T_2+\cdots +T_n$ for $n\in N$ and
$x_0=0, x_n=X_1+X_2+\cdots+X_n, x(t)=x_n$ for $t_n\leq t< t_{n+1}$,
we get a directed CTRW process, where the joint PDF $\phi(x,t)$ can be expressed by $\phi(x,t)=\varphi(t)\delta(x-t)$.
A physiological explanation can be made as follows: the walker has a random time for a rest to
supplement energy, then makes a jump. The longer the rest time is, the jump length can be longer.

Since the variable $x$ takes positive values in proposed directed CTRW model, it is convenient
to replace the Fourier transform for variable $x$ in the formula (5) by the Laplace transform
(i.e. $\widetilde{f}(k)={\cal L}\{f(x);k\}=\int_0^{+\infty}f(x)e^{-kx}dx$) to obtain the following
Laplace-Laplace relation [26]:
\begin{equation}
\label{(6)}
\widetilde{\widetilde{P}}(k,s)=\frac{1-\widetilde{\varphi}(s)}{s}\cdot\frac{1}{1-\widetilde{\widetilde{\phi}}(k,s)}.
\end{equation}
Since
\begin{equation}
\label{(7)}
\begin{array}{lll}
\widetilde{\widetilde{\phi}}(k,s)
&=& \int_0^{+\infty}dt\int_0^{+\infty}\phi(x,t)e^{-kx-st}dx \\
\\
&=& \int_0^{+\infty}dt\int_0^{+\infty}\varphi(t)\delta(x-t)e^{-kx-st}dx\\
\\
&=& \int_0^{+\infty}\varphi(t)e^{-(s+k)t}dt \\
\\
&=& \widetilde{\varphi}(s+k),
\end{array}
\end{equation}
Eq.(6) is recast into
\begin{equation}
\label{(8)}
\widetilde{\widetilde{P}}(k,s)=\frac{1-\widetilde{\varphi}(s)}{s}\cdot\frac{1}{1-\widetilde{\varphi}(s+k)}.
\end{equation}

The $n$th ($n=1,2$) moment of $P(x,t)$ is given by
\begin{equation}
\label{(9)}
\begin{array}{lll}
\langle x^n \rangle (t)
&=&\int_0^{+\infty} x^n(t)P(x,t)dx \\
\\
&=& (-1)^n \frac{\partial^n}{\partial k^n}\widetilde{P}(k,t)\mid_{k=0}\\
\\
&=&  {\cal L}^{-1}\{\frac{1-\widetilde{\varphi}(s)}{s}\cdot (-1)^n\frac{\partial^n}{\partial k^n}\frac{1}{1-\widetilde{\varphi}(s+k)}\mid_{k=0}\}.
\end{array}
\end{equation}

In the following section, we will study the possible behaviors of $P(x,t)$ and its $n$th ($n=1,2$) moment.

\section*{4. The limit distributions of the coupled directed CTRW model}

\ \ \ \ From Eq.(8), we can see that the Laplace-Laplace transform of PDF $P(x,t)$
is completely determined by the Laplace transform of the waiting time PDF $\varphi(t)$.
Usually, the random waiting time is characterized by its mean value $T$.
It may be finite or infinite.

For finite mean waiting time $T$, the Laplace transform of $\varphi(t)$ is of the form
\begin{equation}
\label{(10)}
\widetilde{\varphi}(s)=1-sT+o(s), \hspace{0.5cm} s\rightarrow 0.
\end{equation}

Substituting Eq.(10) into Eq.(8), in the limit $(k,s)\rightarrow (0,0)$, we get the asymptotic relation
\begin{equation}
\label{(11)}
\widetilde{\widetilde{P}}(k,s)\sim\frac{1-(1-sT)}{s}\cdot\frac{1}{1-(1-(s+k)T)}
=\frac{1}{s+k}.
\end{equation}

After taking the inverse Laplace transforms for Eq.(11) about $k$ and $s$, we have
\begin{equation}
\label{(12)}
P(x,t)=\delta(x-t).
\end{equation}

For long times
\begin{equation}
\label{(13)}
\langle x\rangle (t)=t,
\end{equation}
\begin{equation}
\label{(14)}
\langle x^2\rangle (t)=t^2.
\end{equation}

From Eq.(11), we get
\begin{equation}
\label{(15)}
s\widetilde{\widetilde{P}}(k,s)-1+k\widetilde{\widetilde{P}}(k,s)=0.
\end{equation}

Using ${\cal L}\{\frac{\partial P(x,t)}{\partial t};s\}=s\widetilde{P}(x,s)-P(x,0)$,
${\cal L}\{\frac{\partial P(x,t)}{\partial x};k\}=k\widetilde{P}(k,t)-P(0,t)$,
initial condition $P(x,0)=\delta(x)$ and natural boundary conditions, we obtain the
partial differential equation
\begin{equation}
\label{(16)}
\frac{\partial P(x,t)}{\partial t}+ \frac{\partial P(x,t)}{\partial x}=0,
\end{equation}
which is the standard advection equation.

In many applications, one needs to consider a long waiting time (i.e. $T$ is infinite),
it is natural to generalize Eq.(10) to the following form:
\begin{equation}
\label{(17)}
\widetilde{\varphi}(s)=1-s^\beta+o(s^\beta),\hspace{0.5cm} s\rightarrow 0, 0<\beta\leq 1.
\end{equation}

Inserting Eq.(17) into Eq.(8), in the limit $(k,s)\rightarrow (0,0)$, we get the asymptotic relation
\begin{equation}
\label{(18)}
\widetilde{\widetilde{P}}(k,s)\sim\frac{1-(1-s^\beta)}{s}\cdot\frac{1}{1-(1-(s+k)^\beta)}
=\frac{s^{\beta-1}}{(s+k)^\beta}.
\end{equation}

After taking the Laplace inverse transform for Eq.(18) about $s$, one has
\begin{equation}
\label{(19)}
\begin{array}{lll}
\widetilde{P}(k,t)
&=& \frac{t^{-\beta}}{\Gamma(1-\beta)}\ast [e^{-kt}\frac{t^{\beta-1}}{\Gamma(\beta)}]\\
\\
&=& \int_0^{t} e^{-k\tau}\frac{\tau^{\beta-1}(t-\tau)^{-\beta}}{\Gamma(\beta)\Gamma(1-\beta)}d\tau,
\end{array}
\end{equation}
where we use the formulas ${\cal L}\{t^{\beta-1};s\}=\frac{\Gamma(\beta)}{s^\beta}$ for $\beta>0$,
${\cal L}\{e^{-at}g(t);s\}=\widetilde{g}(s+a)$ and ${\cal L}\{(f\ast g)(t);s\}=\widetilde{f}(s)\widetilde{g}(s)$.

According to the formula (9) and Eq.(19), for long times, one gets
\begin{equation}
\label{(20)}
\langle x\rangle (t)=\beta t,
\end{equation}
\begin{equation}
\label{(21)}
\langle x^2\rangle (t)=\frac{\beta(\beta+1)}{2}t^2.
\end{equation}

Then taking the Laplace inverse transform for Eq.(19) about $k$, the following form is obtained
\begin{equation}
\label{(22)}
\begin{array}{lll}
P(x,t) &=& \int_0^{t}\delta(x-\tau)\frac{\tau^{\beta-1}(t-\tau)^{-\beta}}{\Gamma(\beta)\Gamma(1-\beta)}d\tau \\
\\
&=& \frac{x^{\beta-1}(t-x)^{-\beta}}{\Gamma(\beta)\Gamma(1-\beta)},
\end{array}
\end{equation}
which is the density of a random variable $tB$, where $B$ has a
Beta distribution with parameters $\beta$ and $1-\beta$.

From Eq.(19), we can also obtain
\begin{equation}
\label{(23)}
(s+k)^\beta \widetilde{\widetilde{P}}(k,s)=s^{\beta-1},
\end{equation}
which leads to the pseudo-differential equation [27-28]
\begin{equation}
\label{(24)}
(\frac{\partial}{\partial t}+ \frac{\partial}{\partial x})^\beta P(x,t)=\delta(x)\frac{t^{-\beta}}{\Gamma(1-\beta)}
\end{equation}
with a coupled space-time fractional derivative operator on the left-hand side.

Eq.(24) is useful to model flow in porous media and other physical systems characterized
by a link between the waiting time and the jump length.

\section*{5. Conclusions}

\ \ \ \ In this work, we introduce a directed CTRW model with jump
lengths depending on waiting times. By the Laplace-Laplace
transform technique, we find that the PDF $P(x,t)$ is determined
only by the waiting times PDF $\varphi(t)$. For finite and
infinite mean waiting time, we deduce the limit distributions of
$P(x,t)$ from the asymptotic behaviors of $\varphi(t)$ in the
Laplace domain respectively. The corresponding evolving equations
are also derived. For finite mean waiting time, the limit behavior
of the PDF $P(x,t)$ is governed by a standard advection equation.
For infinite mean waiting time, the limit behavior of the PDF
$P(x,t)$ is governed by a pseudo-differential equation with
coupled space-time fractional derivative. We also calculate the
first order moment $\langle x\rangle(t)$ and the second order
moment $\langle x^2\rangle(t)$ of $P(x,t)$. An interesting
phenomenon is obtained: there exist the relations $\langle
x\rangle(t)\sim t$, $\langle x^2\rangle(t)\sim t^2$, whether the
mean waiting time is finite or not.

\section*{Acknowledgements}
\ \ \ \ This project was supported by the Natural Science
Foundation of China (Grant Nos. 11371016 and 11271311), the
Chinese Program for Changjiang Scholars and Innovative Research
Team in University (PCSIRT) (Grant No. IRT1179), the Research
Foundation of Education Commission of Hunan Province of China
(grant no. 11A122), the Lotus Scholars Program of Hunan province
of China.

\end{document}